\documentclass[journal]{IEEEtran}

\pdfoutput=1

\usepackage{amsmath,bm,amssymb,amsfonts,amsthm}
\usepackage[linesnumbered, ruled, vlined]{algorithm2e}
\usepackage{graphicx}
\usepackage{xcolor} 
\usepackage{relsize}
\usepackage{textcomp}
\usepackage{listings}
\usepackage{gensymb}

\usepackage{changepage}
\usepackage{hhline}
\usepackage{multirow}
\usepackage{nomencl}
\usepackage{mathtools}
\usepackage{commath}
\usepackage{booktabs}
\usepackage{subfiles}
\usepackage{tabularx}
\usepackage{lscape}
\usepackage{rotating}

\usepackage{verbatim}
\usepackage{comment}

\usepackage[normalem]{ulem}
\usepackage[utf8]{inputenc}
\usepackage[hyphens]{url}
\usepackage{longtable}
\usepackage{lineno}
    \modulolinenumbers[5]
\usepackage[caption=false]{subfig}
\usepackage{wrapfig}
\usepackage{enumitem}
    \setlist{nolistsep}
\usepackage[hidelinks]{hyperref}
\usepackage{makecell}
\usepackage{multicol}
\usepackage{colortbl}

\usepackage{scalerel}
\usepackage{tikz}
\usetikzlibrary{svg.path}
\definecolor{orcidlogocol}{HTML}{A6CE39}
\tikzset{
  orcidlogo/.pic={
    \fill[orcidlogocol] svg{M256,128c0,70.7-57.3,128-128,128C57.3,256,0,198.7,0,128C0,57.3,57.3,0,128,0C198.7,0,256,57.3,256,128z};
    \fill[white] svg{M86.3,186.2H70.9V79.1h15.4v48.4V186.2z}
                 svg{M108.9,79.1h41.6c39.6,0,57,28.3,57,53.6c0,27.5-21.5,53.6-56.8,53.6h-41.8V79.1z M124.3,172.4h24.5c34.9,0,42.9-26.5,42.9-39.7c0-21.5-13.7-39.7-43.7-39.7h-23.7V172.4z}
                 svg{M88.7,56.8c0,5.5-4.5,10.1-10.1,10.1c-5.6,0-10.1-4.6-10.1-10.1c0-5.6,4.5-10.1,10.1-10.1C84.2,46.7,88.7,51.3,88.7,56.8z};
  }
}
\newcommand\orcidicon[1]{\href{https://orcid.org/#1}{\mbox{\scalerel*{
\begin{tikzpicture}[yscale=-1,transform shape]
\pic{orcidlogo};
\end{tikzpicture}
}{|}}}}

\usepackage[noadjust]{cite}

\begin{document}

\title{\huge Dynamic State Estimation-Based Protection \\ for Induction Motor Loads}

\author{
    Arthur~K.~Barnes $^{1}$\orcidicon{0000-0001-9718-3197},
    Sarbajit~Basu $^{2}$\orcidicon{0000-0002-2581-3089},
    and Adam~Mate $^{1}$\orcidicon{0000-0002-5628-6509}
    \vspace{-0.15in}

\thanks{Manuscript submitted:~Jul.~15,~2022.
Current version: Aug.~15,~2022.
}

\thanks{$^{1}$ The authors are with the Advanced Network Science Initiative at Los Alamos National Laboratory, Los Alamos, NM 87545 USA. \\ Email: abarnes@lanl.gov, amate@lanl.gov.}

\thanks{$^{2}$ The author is with the Klipsch School of Electrical and Computer Engineering at the New Mexico State University, Las Cruces, NM 88003 USA. Email: sarbasu@nmsu.edu.}

\thanks{LA-UR-22-27020. Approved for public release; distribution is unlimited.}

\thanks{978-1-6654-9921-7/22/\$31.00 \copyright2022 IEEE}

}

\markboth{IEEE/PES 54th North American Power Symposium, October~2022}{}

\maketitle


\begin{abstract}
Ensuring protective device coordination is critical to maintain the resilience and improve the reliability of large microgrids. Inverter-interfaced generation, however, poses significant challenges when designing protection systems.
Traditional time-overcurrent protective devices are unsuitable on account of the lack of fault current. Present industry practice is to force all inverters to shut down during faults, which prevents large microgrids from operating in a resilient and reliable manner.
Dynamic state estimation (DSE) has been proposed for line protection, and more recently for the protection of load buses or downstream radial portions of microgrids. However, only passive loads with series resistive-inductive loads have been tested with DSE, even though the behavior of dynamic loads -- such as induction motors or power electronics -- may differ significantly during faults.
This paper considers the case of applying DSE to protecting a load bus serving a three-phase induction motor. 
\end{abstract}


\section{Introduction} \label{sec:introduction}
\indent

Microgrids have been a key tool to integrate distributed renewable energy resources into the bulk energy system.
When operating under islanded conditions, microgrids are reliant upon not only inverter run energy storage, but renewable resources as well. Although they are reasonably reliable, economical, and low greenhouse gas contributors, inverter fed resources -- commonly photovoltaic panels and wind farms -- suffer from issues such as availability -- variability in the levels of irradiation and wind available during the day -- that makes their power output variable and limited system inertia \cite{8076896}.
Under faulted conditions, the current drawn by the system when fed by a synchronous machine is significantly higher and is thus ideal for an overcurrent relay to provide protection. However, an inverter is limited at the amount of current it can provide, therefore limiting the ability to use conventional devices to provide protection; also, the response of an inverter fed device to a change is fast, and thus control devices must respond quickly as well \cite{Aree2021}.

Electrical loads today are dominated by power electronic converter run devices, and single-phase and three-phase induction machines. Most devices have an inverter involved for powering them, while induction motors -- the mainstay of HVAC systems, agricultural systems, and several industrial applications owing to the low cost of construction, easy maintenance, and high efficiency -- require protection mechanisms in place for safe and reliable operation \cite{krause13, en15103564}.
Induction motors, owing to their non-linear behavior, undergo prolonged periods of transiency until they reach steady-state; this warrants the need to ensure that system parameters during the period of transition does not reach extremely high values \cite{brereton57}.
Under faulted conditions -- be it stator faults, rotor faults, or external line faults -- the current drawn by the machine is significantly higher, which in turn can impact the health of the device \cite{8606599}. This becomes more profound in an inverter fed microgrid, where the system has limited to no inertia -- due to the lack of stored kinetic energy -- that could be harvested to adjust to intermittent loading conditions or fault conditions \cite{1325283}.
Consequently, it becomes important to study the protection of induction motors in a microgrid.

Protection methods enable the resilient and reliable operation of microgrids by maintaining protective device coordination to minimize the extent of outages.
Conventional approaches have focused on protecting just the inverters, resulting in the whole microgrid shutting down during a fault; additionally, these approaches do not scale to large networks \cite{9416600}.
Distance relaying techniques have been explored for protecting inverter-interfaced microgrids: Dewadasa et al. \cite{Dewadasa2008} suggested limiting the per phase current in faulted conditions by using distance relaying; Kar et al. \cite{Kar2014} demonstrated the use of differential relaying based on the discrete S-transform performed on the change in system current magnitude under fault conditions; and Huang \cite{Huang2004AdaptiveWA} used an adaptive protection system, in which multiple relays co-ordinate among themselves to predict the occurrence of a fault and recommend corrective action. All these, however, employ the use of legacy programmed protection devices and does not cater to system dynamics or protect the system against bi-directional power flow.
It is unequivocal that new protection methods are needed to support the deployment of large microgrids.

Settingless protection addresses these shortcomings by employing Dynamic State Estimation (DSE) to generate system measurements such that protective control action is determined based on the mismatch between the measurements and estimates generated using accurate system models \cite{Vasios2018, 9416613, Vanin22}.
DSE allows for increased accuracy in tracking system dynamics and better system estimates, therefore allowing better system operation and control \cite{8624411}; it can accurately track system dynamics and can effectively predict the fault characteristics across microgrids of varying sizes under high penetration of renewable energy resources \cite{9429932}.
DSE has been used to develop centralized protection schemes \cite{9429932, choi2017} and to detect hidden failures in substation protection systems \cite{albinali2017b}. More recently, it has been proposed for protection of load buses \cite{Barnes2022-dse-loadbus} and downstream radial sections of microgrids \cite{9416613, dehghanpour2021}. It is also finding an increased relevance in power system operation and monitoring applications \cite{9210125},
owing to the implementation of accurate dynamic models that provide reliable estimates, improved oscillation monitoring, and increased potential for decentralized control \cite{8624411}.

This work investigates the application of DSE for load bus protection, while considering the dynamic behavior of an induction motor.
Section~\ref{sec:methodology} presents the dynamic simulation model selected for an induction motor, the DSE formulation derived from the simulation model, and the solution method for the DSE problem.
Section~\ref{sec:numerical-experiments} presents a case study system based on a numerical simulation of an induction motor under several different fault scenarios.
Section~\ref{sec:results} presents experimental results from the numerical simulation and DSE algorithm.
Finally, Section~\ref{sec:conclusion} summarizes the conclusions regarding the utility of DSE for load bus protection.

\vspace{0.1in}
\section{Methodology} \label{sec:methodology}
\indent

The induction machine model used in this work is a Julia \cite{julia} implementation of the MATLAB asynchronous machine model \cite{matlabAsynchronous}, which has been developed using the induction motor models proposed by \cite{krause13} and \cite{kundur1994power}.

\subsection{5\textsuperscript{th}- order Induction Motor State Estimation Model}

\noindent To convert time-varying induction motor inductances into constant values, in order to simplify the solution of the governing differential equations, the quantities for the machine inputs and states were converted into a rotating reference frame, as described below.

\vspace{0.1in}
\normalsize \noindent Rotating reference-frame transform from \cite{krause13} (3.3-4):
\small
\begin{equation}
\begin{split}
& \begin{bmatrix} v_q(t) \\ v_d(t) \\ v_0(t) \end{bmatrix} = \\
& = \frac{2}{3} \begin{bmatrix} \cos(\theta(t)) & \cos(\theta(t)-\alpha) & \cos(\theta(t)+\alpha) \\ \sin(\theta(t)) & \sin(\theta(t)-\alpha) & \sin(\theta(t)+\alpha) \\ 
1/2 & 1/2 & 1/2 \end{bmatrix} \begin{bmatrix} v_a(t) \\ v_b(t) \\ v_c(t) \end{bmatrix}
\label{eq:dq0-transform}
\end{split}
\end{equation}
\normalsize \noindent where
$\alpha = \frac{2}{3}\pi$;
$v_a(t)$, $v_b(t)$, and $v_c(t)$ are the line-ground voltages on phases A, B and C, respectively; and
$v_q(t)$, $v_d(t)$, and $v_0(t)$ are the voltages on the q, d, and 0 axes, respectively.

\vspace{0.15in}
\normalsize \noindent Magnetizing flux linkages from \cite{krause13} (6.14-15) -- (6.14-17):
\small
\begin{equation}
\dot{\psi}_{qs}(t) = v_{qs}(t) - \omega \cdot \psi_{ds}(t) - R_s \cdot i_{qs}(t)
\label{eq:new-flux1}
\end{equation}
\normalsize \noindent where
$v_{qs}(t)$ is the stator voltage in the q-axis; and
$\psi_{ds}(t)$ is the stator flux in the d-axis.

\small
\begin{equation}
\dot{\psi}_{ds}(t) = v_{ds}(t) + \omega \cdot \psi_{qs}(t) - R_s \cdot i_{ds}(t)
\label{eq:new-flux2}
\end{equation}
\normalsize \noindent where
$v_{ds}(t)$ is the stator voltage in the d-axis; and
$\psi_{qs}(t)$ is the stator flux in the q-axis.

\small
\begin{equation}
\dot{\psi}_{qr}'(t) = v_{qr}'(t) - (\omega-\omega_r) \cdot \psi_{dr}'(t) - R_r' \cdot i_{qr}'(t)
\label{eq:new-flux3}
\end{equation}
\newpage
\normalsize \noindent where
$v_{qr}'(t)$ is the rotor voltage referred to the stator in the q-axis; and
$\psi_{dr}'(t)$ is the rotor flux in the d-axis referred to the stator.

\small
\begin{equation}
\dot{\psi}_{dr}'(t) = v_{dr}'(t) + (\omega-\omega_r) \cdot \psi_{qr}'(t) - R_r' \cdot i_{dr}'(t)
\label{eq:new-flux4} 
\end{equation}
\normalsize \noindent where
$v_{dr}'(t)$ is the rotor voltage referred to the stator in the d-axis; and
$\psi_{qr}'(t)$ is the rotor flux in the q-axis referred to the stator.

\vspace{0.1in}
\normalsize \noindent In Equations (\ref{eq:new-flux1}) -- (\ref{eq:new-flux4}):
$\omega(t)$ is the synchronous speed;
$R_s$ is the stator resistance;
$R'_r$ is the rotor resistance; and
$\omega_r(t) = P \omega(t)$, where
$P$ is the number of poles in the induction machine and
$\omega_r(t)$ is the angular velocity of the rotor.

\vspace{0.1in}
\normalsize \noindent Flux linkages expressed as a function of currents, from \cite{krause13} (6.14-4) -- (6.14-6):
\small
\begin{equation}
{\psi}_{qs}(t) = L_s \cdot i_{qs}(t) + L_m \cdot i_{qr}'(t)
\label{eq:I-flux1}
\end{equation}
\begin{equation}
{\psi}_{ds}(t) = L_s \cdot i_{ds}(t) + L_m \cdot i_{dr}'(t)
\label{eq:I-flux2}
\end{equation}
\begin{equation}
{\psi}_{qs}(t) = L_r' \cdot i_{qr}'(t) + L_m \cdot i_{qs}(t)
\label{eq:I-flux3}
\end{equation}
\begin{equation}
{\psi}_{qs}(t) = L_r' \cdot i_{qr}'(t) + L_m \cdot i_{qs}(t)
\label{eq:I-flux4} 
\end{equation}

\vspace{0.1in}
\normalsize \noindent In Equations (\ref{eq:new-flux1}) -- (\ref{eq:I-flux4}):
$i_{qs}(t)$ and $i_{ds}(t)$ are the stator currents;
$i_{qr}(t)'$ and $i_{dr}(t)'$ are the rotor currents referred to the stator;
$L_s$ is the stator inductance;
$L_r '$ is the rotor inductance; and
$L_m$ is the magnetizing inductance.

\vspace{0.1in}
\normalsize \noindent Electrical torque from \cite{krause13} (6.6-16)--(6.6-17):
\small
\begin{align}
T_e(t) &= \frac{3}{2} \cdot \frac{P}{2} \cdot \frac{1}{\omega} \cdot \left(\psi'_{qr}(t) \cdot i'_{dr}(t) - \psi'_{dr}(t) \cdot i'_{qr}(t) \right)  \\
&= \frac{3}{2} \cdot \frac{P}{2} \cdot \frac{L_m}{D} \frac{1}{\omega} \cdot \left( \psi_{qs}(t) \cdot \psi'_{dr}(t) - \psi'_{qr}(t) \cdot \psi_{ds}(t) \right) 
\label{eq:te}
\end{align}

\vspace{0.1in}
\normalsize \noindent Rotor speed from \cite{krause13} (6.3-8):
\small
\begin{equation}
\dot{\omega}_m(t) = \frac{P}{2J} \cdot \left( T_e(t) - F \cdot w_m - T_m(t) \right)
\label{eq:wr}
\end{equation}
\normalsize \noindent where
$T_e$ is the electromagnetic torque;
$T_m$ is the shaft mechanical torque; 
$J$ is the combined rotor and load inertia coefficient; and
$F$ is the combined rotor and load viscous friction coefficient.

\normalsize
\subsection{5\textsuperscript{th}- order Induction Motor State Estimation Model}

\noindent To perform state estimation, the above equations from the simulation model were reformulated as State-Output Mapping, described in \cite{7981259}.

\vspace{0.1in}
\normalsize \noindent \dashuline{Observables}
\small
\begin{equation}
\mathbf{y} =
\begin{bmatrix}
v_q(t) & v_d(t) & i_q(t) & i_d(t) & \\
z_{\psi_{qs}}(t) & z_{\psi_{ds}}(t) & z_{\psi'_{qr}}(t) & z_{\psi'_ {dr}}(t) & z_{Te}(t) \end{bmatrix}^T
\label{eq:yt}
\end{equation}

\vspace{0.1in}
\normalsize \noindent \dashuline{State}
\small
\begin{equation}
\mathbf{x} =
\begin{bmatrix}
\psi_{qs}(t) & \psi_{ds} & \psi'_{qr}(t) & \psi'_{dr}(t) & T_e(t) & \omega_r(t)
\end{bmatrix}^T
\end{equation}

\newpage
\normalsize \noindent \dashuline{State-Output Mapping}

\vspace{0.1in}
\normalsize \noindent The algebraic voltage equations:
\small
\begin{equation}
\label{eq:VoltageQS}
\begin{split}
v_{qs}(t) = \frac{R_s \cdot L'_{lr}}{D} \cdot \psi_{ds}(t) - \omega(t) \cdot \psi_{qs}(t) - \frac{R_s \cdot L_m}{D} \cdot \psi'_{dr}(t)
\end{split}
\end{equation}
\begin{equation}
\label{eq:VoltageDS}
\begin{split}
v_{ds}(t) = \frac{R_s \cdot L'_{lr}}{D} \cdot \psi_{qs}(t) + \omega(t) \cdot \psi_{ds}(t) - \frac{R_s \cdot L_m}{D} \cdot \psi'_{qr}(t)
\end{split}
\end{equation}

\vspace{0.1in}
\normalsize \noindent Algebraic current equations:
\small
\begin{equation}
\label{eq:CurrentQS}
\begin{split}
i_{qs}(t) = \frac{L'_{lr}}{D} \cdot \psi_{ds}(t) - \frac{L_m}{D} \cdot \psi'_{dr}(t)
\end{split}
\end{equation}
\begin{equation}
\label{CurrentDS}
\begin{split}
i_{qs}(t) = \frac{L'_{lr}}{D} \cdot \psi_{qs}(t) - \frac{L_m}{D} \cdot \psi'_{qr}(t)
\end{split}
\end{equation}
\begin{equation}
\label{eq:CurrentQr}
\begin{split}
i'_{qr}(t) = \frac{1}{L_m} \cdot \left( \psi_{qs}(t) - L_s \cdot i_{qs}(t) \right)
\end{split}
\end{equation}
\begin{equation}
\label{eq:CurrentDr}
\begin{split}
i'_{dr}(t) = \frac{1}{L_m} \cdot \left( \psi_{ds}(t) - L_s \cdot i_{ds}(t) \right)
\end{split}
\end{equation}

\vspace{0.1in}
\normalsize \noindent Algebraic electrical torque equation:
\small
\begin{equation}
\label{eq:electrical-torque}
\begin{split}
z_{te}(t) = T_e(t) - \frac{3}{2} \cdot \frac{P L_m}{D} \cdot \left( \psi_{qs}(t) \cdot \psi'_{qr}(t) - \psi'_{qr}(t) \cdot \psi_{ds}(t) \right)
\end{split}
\end{equation}

\vspace{0.1in}
\normalsize \noindent In Equations (\ref{eq:VoltageQS}) -- (\ref{eq:electrical-torque}):
\small
\begin{equation}
D = L'_{lr} \cdot L_{ls} - L_m^2
\end{equation}

\vspace{0.1in}
\normalsize \noindent Differential flux equations:
\small
\begin{equation}
\begin{split}
\label{eq:z-pqs}
z_{\psi_{qs}}(t) = \psi_{qs}(t) - \psi_{qs}(t-\Delta t) - \int_{t-\Delta t}^t \dot{\psi}_{qs}(\tau) \; d\tau
\end{split}
\end{equation}
\normalsize \noindent where
\small
\begin{equation*}
\begin{split}
\dot{\psi}_{qs}(\tau) = v_{qs}(\tau) - \omega \cdot \psi_{ds}(\tau) - R_s \cdot i_{qs}(\tau)
\end{split}
\end{equation*}
\begin{equation}
\begin{split}
\label{eq:z-pds}
z_{\psi_{ds}}(t) = \psi_{ds}(t) - \psi_{ds}(t-\Delta t) - \int_{t-\Delta t}^t \dot{\psi}_{ds}(\tau) \; d\tau
\end{split}
\end{equation}
\normalsize \noindent where
\small
\begin{equation*}
\begin{split}
\dot{\psi}_{ds}(\tau) = v_{ds}(\tau) + \omega \cdot \psi_{qs}(\tau) - R_s \cdot i_{ds}(\tau)
\end{split}
\end{equation*}
\begin{equation}
\begin{split}
\label{eq:z-pqrp}
z_{\psi'_{qr}}(t) = \psi'_{qr}(t) - \psi_{qr}(t-\Delta t) - \int_{t-\Delta t}^t \dot{\psi}'_{qr}(\tau) \; d\tau
\end{split}
\end{equation}
\normalsize \noindent where
\small
\begin{equation*}
\begin{split}
\dot{\psi}'_{qr}(\tau) = v'_{qr}(\tau) - \left( \omega(\tau) - P \cdot \omega_r(\tau) \right) \cdot \psi'_{dr}(\tau) - R_{rp} \cdot i'_{qr}(\tau)
\end{split}
\end{equation*}
\begin{equation}
\begin{split}
\label{eq:z-pdrp}
z_{\psi'_{dr}}(t) = \psi'_{dr}(t) - \psi_{dr}(t-\Delta t) - \int_{t-\Delta t}^t \dot{\psi}'_{dr}(\tau) \; d\tau
\end{split}
\end{equation}
\normalsize \noindent where
\small
\begin{equation*}
\begin{split}
\dot{\psi}'_{dr}(\tau) = v'_{dr}(\tau) - \left( \omega(\tau) + P \cdot \omega_r(\tau) \right) \cdot \psi'_{qr}(\tau) - R_{rp} \cdot i'_{dr}(\tau)
\end{split}
\end{equation*}

\newpage
\normalsize \noindent Differential rotor speed equation:
\small
\begin{equation}
\begin{split}
\label{eq:z-wm}
z_{\omega_{r}}(t) = \omega_r(t) - \omega_r(t-\Delta t) - \int_{t-\Delta t}^t \dot{\omega}_r(\tau) \; d\tau
\end{split}
\end{equation}
\normalsize \noindent where
\small
\begin{equation*}
\begin{split}
\dot{\omega}_r(\tau) = \frac{P}{2J} \cdot \left( T_e(\tau) - F \cdot \omega_r(\tau) - T_m(\tau) \right)
\end{split}
\end{equation*}
 
\vspace{0.1in}
\normalsize \noindent \dashuline{Discretized Observables}
\small
\begin{equation}
\mathbf{y} =
\begin{bmatrix}
\mathbf{v}_d & \mathbf{v}_q & \mathbf{i}_d & \mathbf{i}_q & \\
\mathbf{z}_{\psi_{qs}} & \mathbf{z}_{\psi_{ds}} & \mathbf{z}_{\psi'_{qr}} & \mathbf{z}_{\psi'_ {dr}} & \mathbf{z}_{Te} \end{bmatrix}^T
\label{eq:yi}
\end{equation}

\vspace{0.1in}
\normalsize \noindent \dashuline{Discretized State}
\small
\begin{equation}
\mathbf{x} =
\begin{bmatrix}
\mathbf{\psi}_{qs} & \psi_{ds} & \mathbf{\psi}'_{qr} & \mathbf{\psi}'_{dr} & \mathbf{T}_e & \mathbf{\omega}_r
\end{bmatrix}^T
\end{equation}

\vspace{0.1in}
\normalsize \indent To perform dynamic state estimation, which was formulated as a discrete-time problem, it is necessary to discretize the governing equations for the induction motor. The discretization of the algebraic equations is trivial, while the discretization of the above differential equations was performed with the trapezoidal rule:

\small
\begin{equation}
f[i] - f[i-1] \approx \frac{1}{\Delta t} \int_{t-\Delta t}^t f(\tau) d\tau
\end{equation}

\normalsize
\subsection{Solution of the State Estimation Model}

\noindent Given the discretized state-output relationship $h(n)$, the Jacobian $H(n,n)$ can be calculated either analytically -- based on the discretized algebraic and diffential state equations -- or numerically.
While for a real-time implementation an algebraic representation of the Jacobian is required, for a proof-of-concept -- such as that illustrated here -- a numerical Jacobian can be calculated. For this study, the Julia library FiniteDiff.jl \cite{julia-finitediff} is used for numerical Jacobian calculation.

\vspace{0.1in}
Given $h(\cdot)$ and $\mathbf{H}$, the state of the system can be solved by the following updated equations:
\small 
\begin{equation}
\epsilon_i = y - h(x_i)
\end{equation}
\begin{equation}
x_{i+1} = x_{i} + (\mathbf{H}_i^T \cdot \mathbf{H}_i)^{-1}\mathbf{H}_i^T \cdot \epsilon_i
\end{equation}

\normalsize \noindent This process is repeated iteratively until either the maximum number of iterations is reached or the algorithm has converged, indicated by the change in the log of the squared error falling below a specified threshold:

\small
\begin{equation}
\Delta J_i = |\log |\epsilon_i^* \cdot \epsilon_i| - \log |\epsilon_{i-1}^* \cdot\epsilon_{i-1}||
\end{equation}

\vspace{0.05in}
\normalsize \indent The measurement error test is performed as follows:
\small
\begin{equation}
p = F_{m-n}(J_i) \ge 0.95
\end{equation}
\normalsize \noindent where $F_{m-n}$ is the Chi-squared cumulative distribution function for $m - n$ degrees of freedom, in which $m - n$ is the number of linearly independent observables.

\vspace{0.1in}
\normalsize \indent The final algorithm is illustrated below:
\newpage
\begin{algorithm}
\caption{DSE Algorithm Pseudocode}
\label{alg:dse-pseudocode}
\DontPrintSemicolon
\SetAlgoLined
\small
    initialize state vector \textit{$x$} to small normally-distributed random numbers \;
    \While{change in error is greater than threshold \textbf{or} maximum iteration limit is reached}{
        calculate the estimated output \textit{$y_{est}$} from h(x) \;
        calculate the sum-squared error \textit{$J$} between measured output \textit{$y$} and \textit{$y_{est}$} \;
        calculate the change in the state variable \textit{$x$} based on the update equation \;
    }
    calculate chi-squared CDF based on the final value of \textit{$J$} \;
\end{algorithm}

\section{Numerical Experiments} \label{sec:numerical-experiments}
\indent

The DSE algorithm was implemented in Julia~1.7 64-bit, on a computer with an Intel Xeon\textsuperscript{\textregistered} E-2276M 2.8 GHz CPU with 128~GB RAM, running MS Windows 10~21H2.

\begin{figure}[!htbp]
\centering
\includegraphics[width=0.45\textwidth]{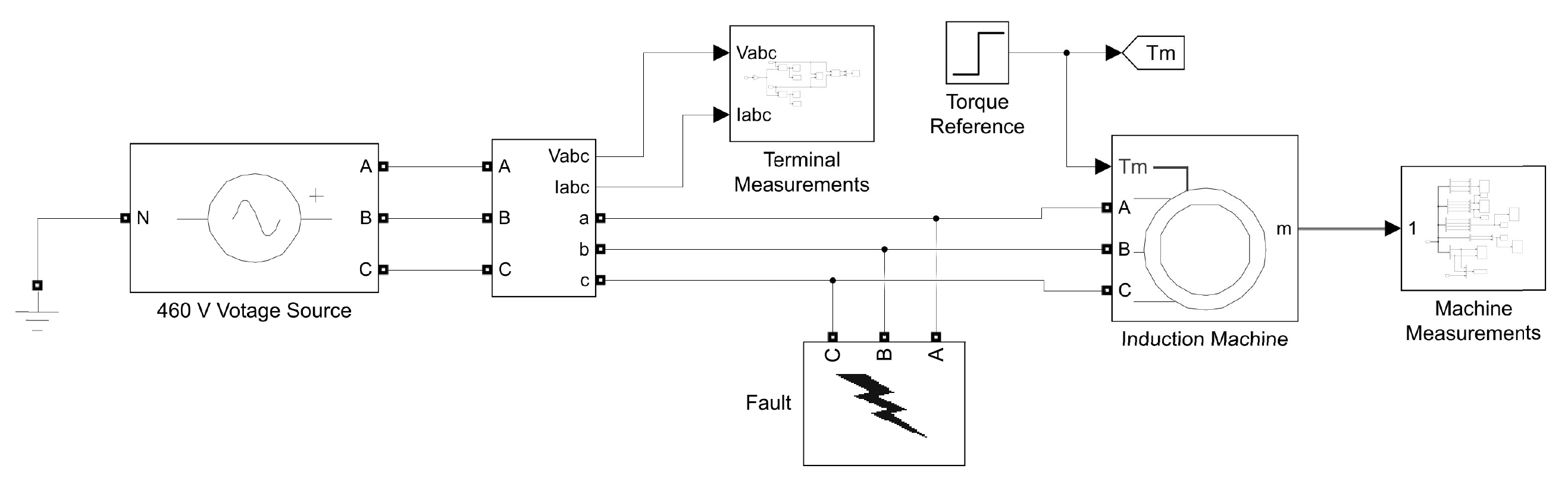}
\caption{MATLAB Simulink\textsuperscript{\textregistered} SimScape model for an induction motor with a fault at the terminals.}
\label{fig:im-fault-model}
\end{figure}

\vspace{0.05in}
The DSE algorithm was applied to a small single-bus case study system, illustrated in Fig.~\ref{fig:im-fault-model}.
It was implemented in MATLAB Simulink\textsuperscript{\textregistered} in the Specialized Power Systems blockset.
The system consists of a 3.75~kW (5 hp) 460~V induction motor, supplied by a voltage source, with parameters illustrated in Table~\ref{table:case-study-params}. The motor is direct-online started and drives a constant-torque load.

\begin{table}[!htbp]
\centering
\caption{Case Study Parameters}
\begin{tabular}{l l l r l} 
\hline \hline
Subsystem & Parameter & Symbol & Value & Units \\ [0.5ex] 
\hline
& Nominal Power & $P$ & 3.73 & kW \\
& Voltage (line-line) & $V_{ll}$ & 460 & V \\
& Frequency & $f$ & 60 & Hz \\
& Pole Pairs & $P$ & 2 & \\
& Inertia & $J$ & 0.02 & $\mathrm{kg\cdot m^2}$ \\
Stator & Resistance & $R_s$ & 1.115 & $\Omega$ \\
Stator & Inductance & $L_{ls}$ & 5.974 & mH \\
Rotor & Resistance & $R_r'$ & 1.083 & $\Omega$ \\
Rotor & Inductance & $L_{lr}'$ & 5.974 & mH \\
Rotor & Mutual Inductance & $L_m$ & 203.7 & mH \\
Load & Torque & $T_m$ & 50 & $\mathrm{N\cdot m}$ \\
Fault & Fault Resistance & $R_f$ & 5.0 & $\Omega$ \\
Fault & Ground Resistance & $R_g$ & 0.1 & $\Omega$ \\
\hline \hline
\end{tabular}
\label{table:case-study-params}
\end{table}

\noindent The sequence of events in the simulation is as follows: 
\begin{enumerate}
\item the motor is connected to power at $0~sec$,
\item the constant-torque load is activated at $3~sec$,
\item a fault is applied at $5~sec$, and
\item the fault is cleared at $5.25~sec$.
\end{enumerate}

\newpage
\noindent The following quantities are recorded during the simulation:
\begin{itemize}
\item line-ground voltage on each phase at the voltage source,
\item current on each phase at the voltage source,
\item mechanical torque drawn by the load, and
\item rotor shaft speed.
\end{itemize}

\vspace{0.05in}
\noindent For the voltage and current measurements, Park's transformation was performed to produce the discretized set of observables in (\ref{eq:yi}).

\vspace{0.05in}
\noindent Four different scenarios were considered (Table~\ref{table:case-study-results}). 
The transformed measurements were down-sampled to a 100~Hz sample rate and applied to the DSE algorithm during the fault period.

\begin{table}[!htbp]
\centering
\caption{Case Study Results}
\begin{tabular}{l l r} 
\hline \hline
Case & Phasing & $\chi^2$ CDF \\ [0.5ex] 
\hline \\
No Fault & -- & 0.988 \\
Line-Ground Fault & AG & 0.925 \\
Line-Line Fault & AB & 0.413 \\
Three-Phase Ground Fault & ABCG & 0.800 \\
\hline \hline
\end{tabular}
\label{table:case-study-results}
\end{table}

\section{Experiment Results} \label{sec:results}
\indent

\begin{figure}[!htbp]
\centering
\includegraphics[width=0.4\textwidth]{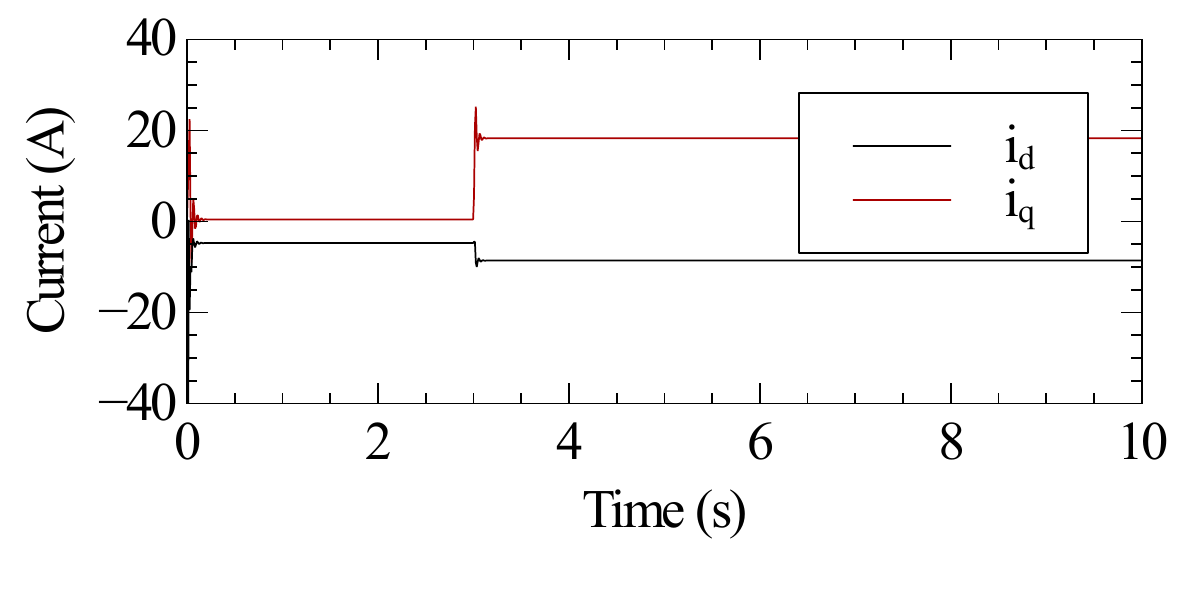}
\caption{Measured voltage and current for unfaulted conditions.}
\label{fig:im-no-fault}
\end{figure}

\begin{figure}[!htbp]
\centering
\includegraphics[width=0.4\textwidth]{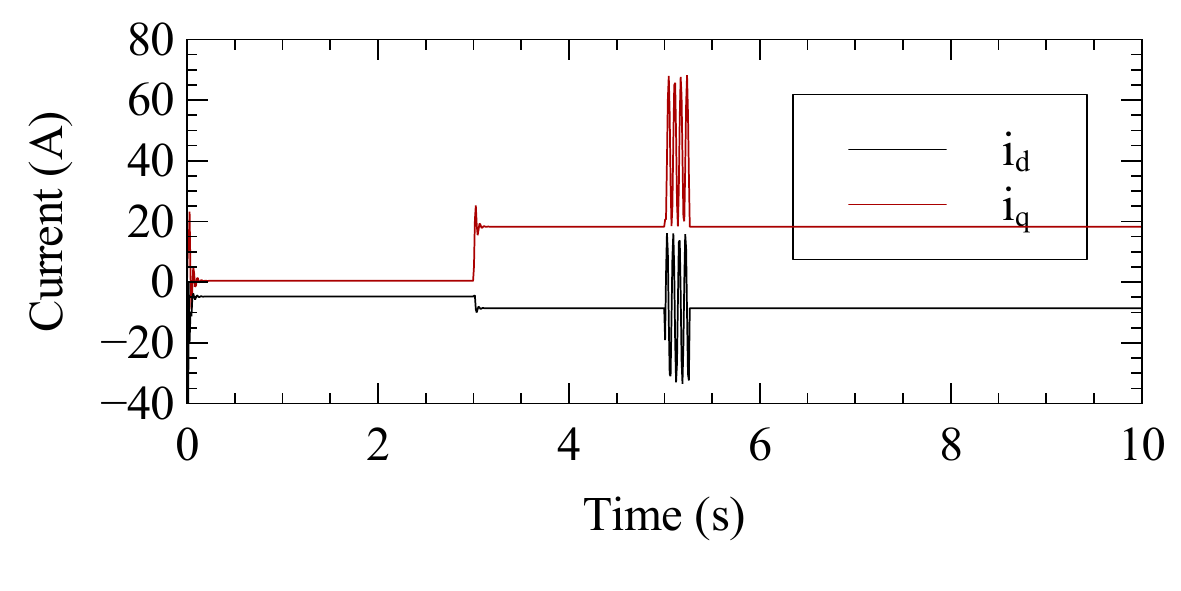}
\caption{Measured voltage and current for a line-ground fault.}
\label{fig:im-lg-fault}
\end{figure}

\begin{figure}[!htbp]
\centering
\includegraphics[width=0.4\textwidth]{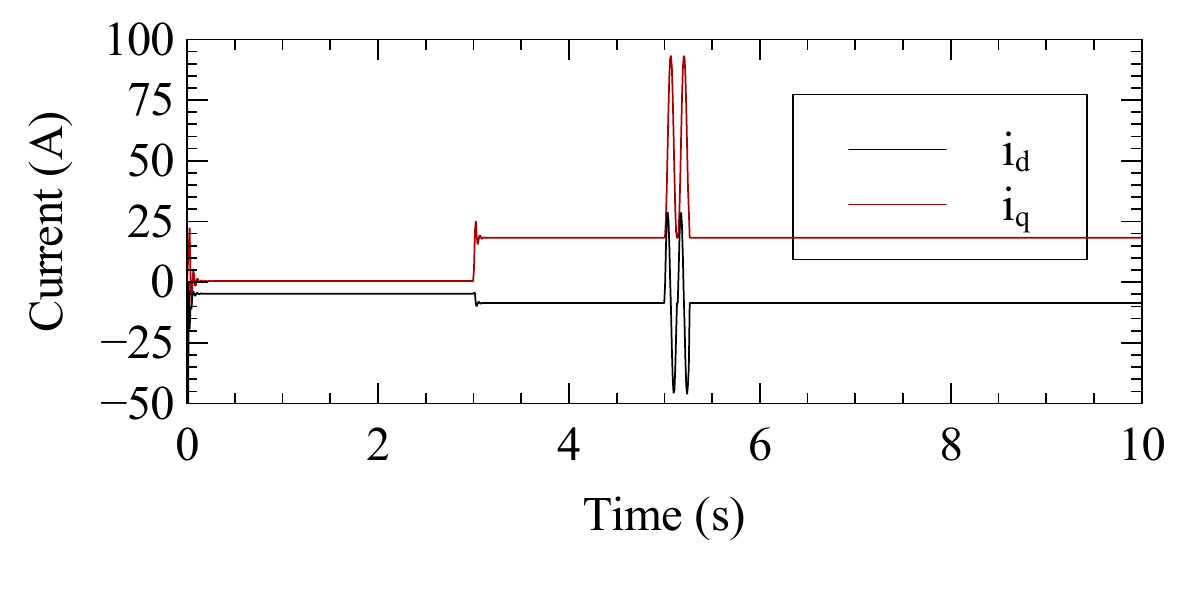}
\caption{Measured voltage and current for a line-line fault.}
\label{fig:im-ll-fault}
\end{figure}

\begin{figure}[!htbp]
\centering
\includegraphics[width=0.4\textwidth]{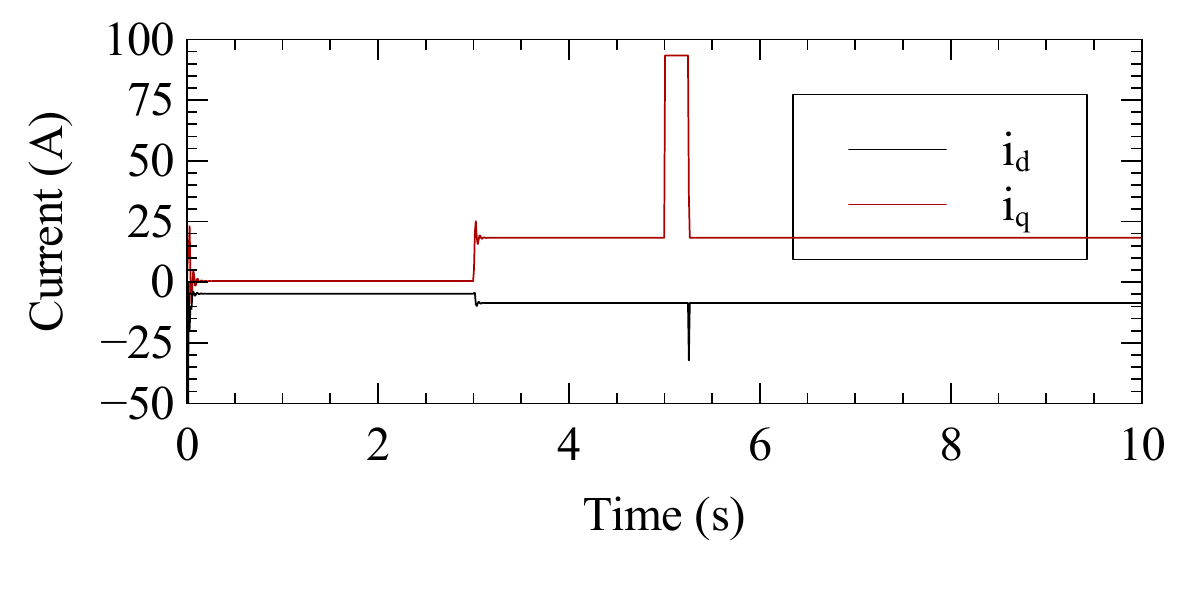}
\caption{Measured voltage and current for a three-phase-ground fault.}
\label{fig:im-3p-fault}
\end{figure}

Measured voltages and currents, for the four fault scenarios considered, are illustrated in Figs.~\ref{fig:im-no-fault} -- \ref{fig:im-3p-fault}.
Applying the DSE algorithm during the fault period yields the Chi-Squared statistic in column~3 of Table~\ref{table:case-study-results}.
Testing to a 95~\% confidence interval allow for faulted conditions to be detected.

\newpage
\section{Conclusions} \label{sec:conclusion}
\indent

This work investigated the application of DSE for load bus protection with dynamic loads, in the case of a three-phase induction motor.
Results suggest that DSE is a feasible method for protection under common fault types based on the ability of the DSE to accurately detect faults on the case study system with a confidence test.

As DSE is a generalisation of differential protection, it can be applied to a wide range of power conversion elements: not only to conventional elements such as transformers, transmission lines, and shunt capacitors, but to elements such as motor loads in microgrids as well. 
It is expected that this protection method can be applied to other forms of dynamic loads, such as single-phase induction motors or active rectifiers as well. Additionally, it is expected that this method could be extended to not only model individual loads, but to be representative of composite load models as well -- along the lines of the WECC CMPLDW model -- representing the dynamic behavior of the downstream radial portions of microgrids.

Future work will aim to investigate these above described other applications, in addition to the feasibility of applying DSE for load bus protection in a real-time implementations.


\bibliographystyle{unsrt}
\bibliography{references}

\end{document}